\documentclass[reqno,twocolumn]{article}
\usepackage{amsmath,amsthm}
\usepackage{graphicx}
\usepackage{hyperref}
\usepackage{epsfig}
\usepackage{psfrag}
\usepackage{graphics}
\begin{document}
\title{
Ensembles of Protein Molecules as Statistical Analog Computers
\footnote{in press}}

\author{Victor Eliashberg \footnote{visit the web site www.brain0.com for information}\\
\\
\emph{Avel Electronics, Palo Alto, California, www.brain0.com}}
\date{}
\maketitle \setlength{\textwidth}{7.0in}
\begin{abstract}
A class of analog computers built from large numbers of
microscopic probabilistic machines is discussed. It is postulated
that such computers are implemented in biological systems
 as ensembles of protein molecules.
The formalism is based on an abstract computational model referred
to as \emph{Protein Molecule Machine (PMM)}. A PMM is a
continuous-time first-order Markov system with real input and
output vectors, a finite set of discrete states, and the
input-dependent conditional probability densities of state
transitions. The output of a PMM is a function of its input and
state. The components of input vector, called \emph{generalized
potentials}, can be interpreted as membrane potential, and
concentrations of neurotransmitters. The components of output
vector, called \emph{generalized currents}, can represent ion
currents, and the flows of second messengers. An \emph{Ensemble of
PMMs (EPMM)} is a set of independent identical PMMs with the same
input vector, and the output vector equal to the sum of output
vectors of individual PMMs. The paper suggests that biological
neurons have much more sophisticated computational resources than
the presently popular models of artificial neurons.

\end{abstract}
\section{Introduction} \label{sec1}

 After the classical work of Hodgkin and Huxley \cite{bib1}, it is widely
 recognized that the conformational changes in the sodium and potassium
 channels account for the generation of nerve spike.  In this specific
 case, the time constants of the corresponding temporal  processes are
 rather small (on the order of a few milliseconds). It is known that
 in  some other cases (such as the ligand-gated  channels) the time constants
 associated  with conformational changes in protein molecules can have much larger
 values \cite{bib4,bib5,bib8}. The growing body of evidence suggests that
 such slower conformational changes have direct behavioral implications
 \cite{bib6, bib7, bib2}. That is, the dynamical computations
 performed by ensembles of protein molecules at the level of individual cells
 play important role in complex neuro-computing processes.

An attempt to formally connect some effects of cellular dynamics
with statistical dynamics of conformations of membrane proteins
was made in \cite{bib3}. The present paper discusses a
generalization of this formalism. The approach is based on an
abstract computational model referred to as \emph{Protein Molecule
Machine (PMM)}. The name expresses the hypothesis that such
microscopic machines are implemented in biological neural networks
as protein molecules. A PMM is a continuous-time first-order
Markov system with real input and output vectors, a finite set of
discrete states, and the input-dependent conditional probability
densities of state transitions. The output of a PMM is a function
of its input and state.

The components of input vector, called \emph{generalized
potentials}, can be interpreted as membrane potential, and
concentrations of neurotransmitters. The components of output
vector, called \emph{generalized currents}, can be viewed as ion
currents, and the flows of second messengers.

An \emph{Ensemble of PMMs (EPMM)} is a set of independent
identical PMMs with the same input vector, and the output vector
equal to the sum of output vectors of individual PMMs. The paper
explains how interacting EPMMs can work as robust statistical
analog computers performing a variety of complex computations at
the level of a single cell.

The EPMM formalism suggests that much more computational resources
are available at the level of a single neuron than is postulated
in traditional computational theories of neural networks. It was
previously shown \cite{bib2, bib10} that such cellular
computational resources are needed for the implementation of
context-sensitive associative memories (CSAM) capable of producing
various effects of working memory and temporal context.

A computer program employing the discussed formalism was
developed. The program, called CHANNELS, allows the user to
simulate the dynamics of a cell with up to ten different
voltage-gated channels, each channel having up to eighteen states.
Two simulation modes are supported: the Monte-Carlo mode (for the
number of molecules from 1 to 10000), and the continuous mode (for
the infinite number of molecules).  New program capable of
handling more complex types of PMMs is under development. (Visit
the web site www.brain0.com for more information.)

\smallskip
\noindent The rest of the paper consists of the following
sections:
\begin{enumerate}
\setcounter{enumi}{1} \item  Abstract Model of Protein Molecule
Machine (PMM) \item Example: Voltage-Gated Ion Channel as a PMM
\item Abstract Model of Ensemble of Protein Molecule Machines
(EPMM) \item EPMM as a Robust Analog Computer \item Replacing
Connections with Probabilities \item Examples of Computer
Simulation \item EPMM as a Distributed State Machine (DSM) \item
Why Does the Brain Need Statistical Molecular Computations? \item
Summary
\end{enumerate}

\section{Abstract Model of Protein Molecule Machine (PMM) } \label{sec2}
A \emph{Protein Molecule Machine} (PMM) is an abstract
probabilistic computing system
$(\textbf{X},\textbf{Y},\textbf{S},\alpha,\omega)$, where
\begin{itemize}
\item \textbf{X} and \textbf{Y} are the sets of real input and
output vectors, respectively  \item \textbf{S}$=\{
s_{0},..s_{n-1}\}$ is a finite set of states  \item $\alpha:
\textbf{X}\times\textbf{S}\times\textbf{S} \to \textbf{R}^\prime$
is a function describing the input-dependent conditional
probability densities of state transitions, where
$\alpha(x,s_i,s_j)dt$ is the conditional probability of transfer
from state $s_j$ to state $s_i$ during time interval $dt$, where
$x\in \textbf{X}$ is the value of input, and $\textbf{R}^\prime$
is the set of non-negative real numbers. The components of $x$ are
called \emph{generalized potentials}. They can be interpreted as
membrane potential, and concentrations of different
neurotransmitters. \item $\omega:
\textbf{X}\times\textbf{S}\rightarrow\textbf{Y}$ is a function
describing output. The components of $y$ are called
\emph{generalized currents}. They can be interpreted as ion
currents, and the flows of second messengers.
\end{itemize}
\noindent Let $x\in\textbf{X}$, $y\in\textbf{Y}$, $s\in\textbf{S}$
be, respectively, the values of input, output, and state at time
t, and  let $P_{i}$ be the probability that $s=s_i$. The work of a
PMM is described as follows:

\begin{align}
   &\frac{dP_i}{dt} = \sum_{j\neq i}{\alpha(x,s_i,s_j)P_j } - P_i \sum_{j\neq i}{\alpha(x,s_j,s_i)}
   \label{eq1}\\
   &at\  t=0 \qquad \sum_{i=0}^{n-1}{P_i} = 1 \label{eq2}\\
   &y =\omega(x,s) \label{eq3}
\end{align}
Summing the right and the left parts of (\ref{eq1}) over
$i=~0,..n-1$ yields
\begin{equation} \label{eq4}
    \frac{d (\sum_{i=0}^{n-1}{P_i })}{dt}= 0
\end{equation}
so the condition (\ref{eq2}) holds for any t.
\begin{figure}[t!]
\begin{center}
 \includegraphics[width=2.6in]{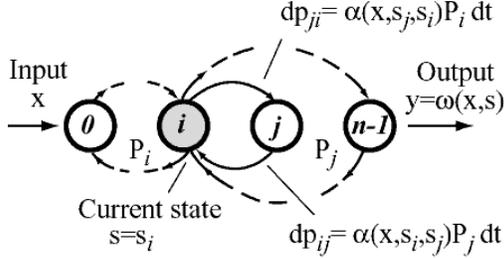}
 \end{center}
 \caption{Internal structure of PMM} \label{fi1}
\end{figure}

\noindent The internal structure of a PMM is shown in Figure~1,
where $dp_{ij}$ is the probability of transition from state
$s_{j}$ to state $s_{i}$ during time interval $dt$. The gray
circle indicates the current state $s=s_i$. The output
$y=~\omega(x,s)$ is a function of input and the current state.

\noindent For the probability of transition from state $s_j$ to
state $s_i$ we have
\begin{equation} \label{eq5}
   dp_{ij} = \alpha(x,s_i,s_j) P_j dt
\end{equation}
It follows from (\ref{eq1}) that
\begin{equation} \label{eq6}
  dP_i = \sum_{j\neq i}{(dp_{ij}-dp_{ji})}
\end{equation}
\section{Example: Voltage-Gated Ion Channel as a PMM} \label{sec3}
Ion channels are studied by many different disciplines:
biophysics, protein chemistry, molecular genetics, cell biology
and others (see extensive bibliography in \cite{bib5}).
\begin{figure}[b!]
\begin{center}
 \includegraphics[width=3.2in]{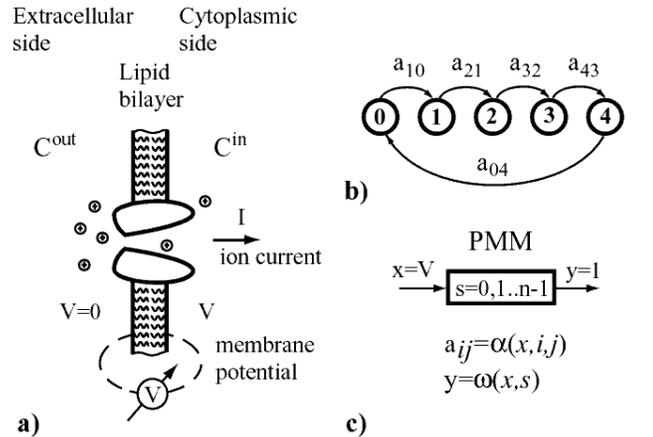}
 \end{center}
 \caption{Ion channel as a PMM} \label{fi2}
\end{figure}
This paper is concerned with the information processing
(computational) possibilities of ion channels.

I postulate that, at the information processing level, ion
channels (as well as some other membrane proteins) can be treated
as PMMs. That is, at this level, the exact biophysical and
biochemical mechanisms are not important. What is important are
the properties of ion channels as abstract machines.

This situation can be meaningfully compared with the general
relationship between statistical physics and thermodynamics. Only
some properties of molecules of a gas (e.g., the number of degrees
of freedom) are important at the level of thermodynamics.
Similarly, only some properties of protein molecules are important
at the level of statistical computations implemented by the
ensembles of such molecules.

The general structure of a voltage-gated ion channel is shown
schematically in Figure \ref{fi2}a. Figures \ref{fi2}b and
\ref{fi2}c show how this channel can be represented as a PMM. In
this example the PMM has five states $s\in \{0,1,..4\}$, a single
input $x=V$ (the membrane potential) and a single output $y=I$
(the ion current). Using the Goldman-Hodgkin-Katz (GHK) current
equation we have the following expression for the output function
$\omega(x,s)$.
\begin{equation} \label{eq7}
   I_j=\omega(V,j)=\frac{p_jz^2FV^{\prime}(C^{in}-C^{out}e^{-zV^\prime}) }{1-e^{-zV^\prime}}
\end{equation}

where
\begin{itemize}
\item $I_j$ is the ion current in state $s=j$ with input $x=V$
\item $p_j\ [cm/sec]$ is the permeability of the channel in state
$s=j$ \item $z$ is the valence of the ion ($z=1$ for $K^+$ and
$Na^+$, $z=2$ for $Ca^{++}$) \item $F=9.6484\cdot10^4\ [C/mol]$ is
the Faraday constant \item $V^{\prime}=\frac{VF}{RT}$ is the ratio
of membrane potential to the thermodynamic potential, where $T\
[K]$ is the absolute temperature, $R=8.3144\ [J/K \cdot~mol]$ is
the gas constant \item $C^{in}$ and $C^{out}\ [mol]$ are the
cytoplasmic and extracellular concentrations of the ion,
respectively
\end{itemize}
One can make different assumptions about the function
$\alpha(x,s_j,s_i)$, describing the conditional probability
densities of state transitions. It is convenient to represent this
function as a matrix of voltage dependent coefficients
$a_{ij}(V)$.
\begin{equation} \label{eq9}
   \alpha =
   \begin{pmatrix}
                 &a_{00}(V)\ ..\ a_{0j}(V)\ ..\  a_{0m}(V)\\
                 \\
                 &a_{i0}(V)\ ..\ a_{ij}(V)\ ..\  a_{im}(V)\\
                 \\
                 &a_{m0}(V)\ ..\ a_{mj}(V)\ ..\  a_{mm}(V)
   \end{pmatrix}
\end{equation}
where $m=n-1$. Note that the diagonal elements of this matrix are
not used in equation (\ref{eq1}).

In the model of spike generation discussed in \cite{bib3}  both
sodium, $Na^+$, and potassium, $K^+$ channels were treated as PMMs
with five states shown in Figure~\ref{fi2}. Coefficients $a_{10}$,
$a_{21}$, $a_{32}$ where assumed to be sigmoid functions of
membrane potential, and coefficients $a_{43}$ and $a_{04}$ -
constant. In the case of the sodium channel, $s=3$ was used as a
high permeability state, and $s=4$ was used as inactive state. In
the case of potassium channel, $s=3$ and $s=4$ were assumed to  be
high permeability states.

\medskip \noindent \textbf{Note}. As the experiments with the
program CHANNELS (mentioned in Section \ref{sec1}) show, in a
model with two voltage-gated channels ($K^+$ and $Na^+$), the
spike can be generated with many different assumptions about
functions $\alpha$ and $\omega$.
 \section{Abstract Model of Ensemble of Protein Molecule Machines
 (EPMM)} \label{sec4}
\begin{figure}[b!]
\begin{center}
\includegraphics[width=2.0in]{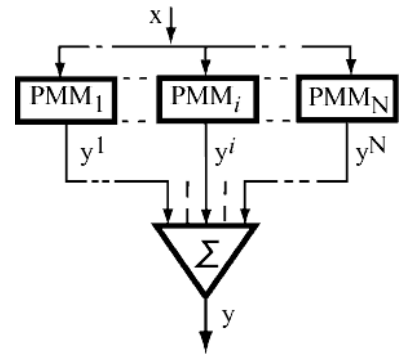}
\end{center}
\caption{The structure of EPMM } \label{fi3}
\end{figure}
An \emph{Ensemble of Protein Molecule Machines} (EPMM) is a set of
identical independent PMMs with the same input vector, and the
output vector equal to the sum of output vectors of individual
PMMs.

The structure of an EPMM is shown in Figure \ref{fi3}, where $N$
is the total number of PMMs, $y^k$ is the output vector of the
k-th PMM, and $y$ is the output vector of the EPMM. We have

\begin{equation} \label{eq10}
   y=\sum_{k=1}^{N}y^k
\end{equation}

Let $N_i$ denote the number of PMMs in state $s=i$ (the occupation
number of state $i$). Instead of (\ref{eq10}) we can write
\begin{equation} \label{eq11}
   y = \sum_{i=0}^{n-1}{N_i\omega(x,s_i)}
\end{equation}

\noindent $N_i$ $(i=0,...n-1)$ are random variables with the
binomial probability distributions
\begin{equation} \label{eq12}
   P \{ N_i=m \}=\binom{m}{N}P_i^m(1-P_i)^{N-m}
\end{equation}
$N_i$ has the mean $\mu_i=NP_i$ and the variance
$\sigma_i^2=~NP_i(1-P_i)$.

\medskip
\noindent Let us define the relative number of PMMs in state $s=i$
(the relative occupation number of state $i$) as
\begin{equation} \label{eq13}
   e_i=\frac{N_i}{N}
\end{equation}

The behavior of the average $\overline{e}_i$ is described by the
equations similar to (\ref{eq1}) and~(\ref{eq2}).
\begin{align}
   &\frac{d\overline{e}_i}{dt} = \sum_{j\neq i}{\alpha(x,s_i,s_j)\overline{e}_j } - \overline{e}_i \sum_{j\neq i}{\alpha(x,s_j,s_i)}
   \label{eq14}\\
   &at\  t=0 \qquad \sum_{i=0}^{n-1}{\overline{e}_i}=1
   \label{eq15}
   \end{align}
\noindent The average output $\overline{y}$ is equal to the sum of
average outputs for all states.

\begin{equation} \label{eq16}
   \overline{y}= N\sum_{i=0}^{n-1}{\omega(x,s_i)\overline{e}_i}
\end{equation}
The standard deviation for $e_k$ is equal to
\begin{equation} \label{eq17}
   \sigma_k = \sqrt{P_k(1-P_k)/N}
\end{equation}
\noindent It is convenient to think of the relative occupation
numbers $e_k$ as the states of analog memory of an EPMM. In
\cite{bib2,bib10,bib3} the  states of such dynamical cellular
short-term memory (STM) were called \emph{E-states}.
\begin{figure}[b!]
\begin{center}
 \includegraphics[width=3.0in]{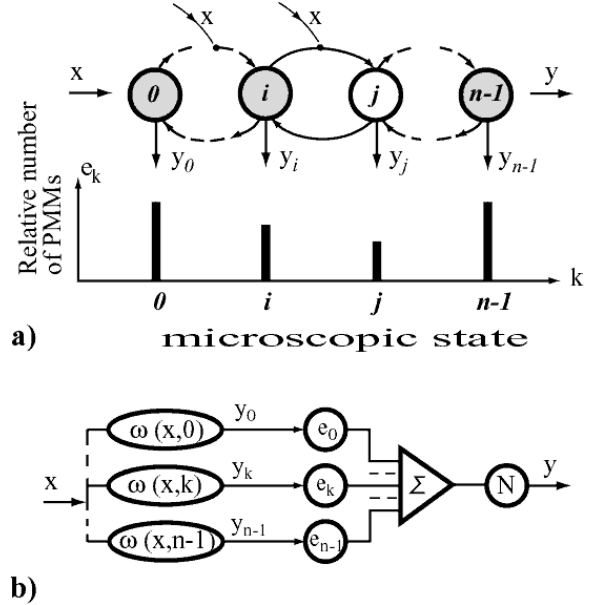}
 \end{center}
 \caption{Relative occupation numbers of the microscopic states of a PMM
 as the macroscopic states of analog memory of the corresponding EPMM } \label{fi4}
\end{figure}

\smallskip
Figure \ref{fi4} illustrates the implementation of E-states as
relative occupation numbers of the microscopic states of a PMM.
The number of independent E-state variables is equal to $n-1$. The
number is reduced by one because of the additional equation
(\ref{eq15}).

\section{EPMM as a Robust Analog Computer} \label{sec5}

\begin{figure}[b!]
\begin{center}
 \includegraphics[width=3.1in]{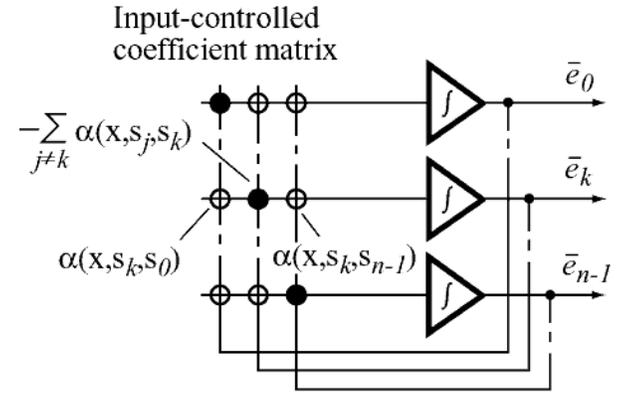}
 \end{center}
 \caption{Simulation of an EPMM using integrating operational amplifiers
 with input--controlled coefficient matrix}\label{fi5}
\end{figure}

An EPMM can serve as a robust analog  computer with the
input--controlled coefficient matrix shown in Figure \ref{fi5}.
 Since all the characteristics of the
statistical implementation of this computer are determined by the
properties of the underlying PMM, this statistical molecular
implementation is very robust.

The implementation using $n$ integrating operational amplifiers
shown in Figure \ref{fi5} is not very reliable. The integrators
based on operational amplifiers with negative capacitive feedback
are not precise, so condition (\ref{eq15}) will be gradually
violated. (A better implementation should use any $n-1$ equations
from (\ref{eq14}) combined with equation $(\ref{eq15})$.) In the
case of the discussed statistical implementation condition
$(\ref{eq15})$ is guaranteed because the number of PMMs, $N$, is
constant.

\section{Replacing Connections with Probabilities} \label{sec6}
\begin{figure}[b!]
\begin{center}
 \includegraphics[width=2.8in]{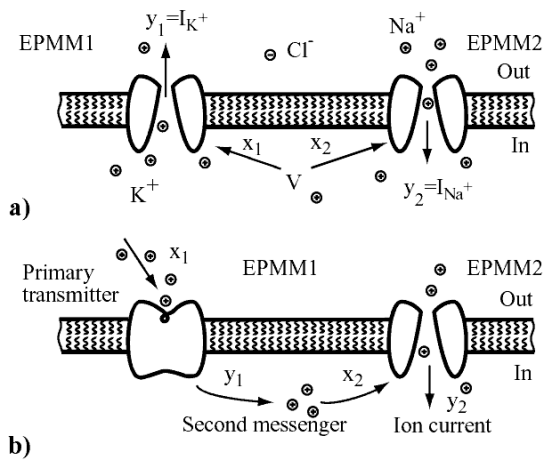}
 \end{center}
 \caption{Two EPMMs interacting via a) electrical and b) chemical messages}\label{fi6}
\end{figure}
The most remarkable property of the statistical implementation of
the analog computer shown in Figure \ref{fi5} is that the matrix
of input-dependent macroscopic connections is implemented as the
matrix of input-dependent microscopic probabilities. For a
sufficiently large number of states (say, $n>10$), it would be
practically impossible to implement the corresponding analog
computers (with required biological dimensions) relying on
traditional electronic operational amplifiers with negative
capacitive feedbacks that would have to be connected via difficult
to make matrices of input-dependent coefficients.

 A single neuron can have many different EPMMs interacting via
 electrical messages (membrane potential) and chemical messages
 (different kinds of neurotransmitters). As  mentioned in
 Section \ref{sec3}, the Hodgkin-Huxley \cite{bib1} model can be naturally
expressed in terms of two  EPMMs (corresponding to the sodium and
potassium channels) interacting via common membrane potential (see
Figure \ref{fi6}a).  Figure \ref{fi6}b shows two EPMMs interacting
via a second messenger. In this example, EPMM1 is the primary
transmitter receptor and EPMM2 is the second messenger receptor.

\section{Examples of Computer Simulation } \label{sec7}
\begin{figure}[b!]
\begin{center}
 \includegraphics[width=2.1in]{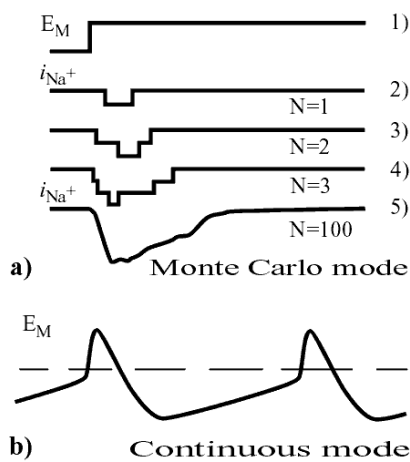}
 \end{center}
 \caption{Examples of computer simulation by program
 CHANNELS}\label{fi7}
\end{figure}
\noindent Figure \ref{fi7} presents examples of computer
simulation done by program CHANNELS mentioned in
Section~\ref{sec1}. Lines 2-4 in Figure \ref{fi7}a display  random
pulses of sodium current produced  by 1, 2, and 3 PMMs,
respectively, representing sodium channel, in response to the
pulse of membrane potential shown in line 1. Line 4 shows a
response of 100 PMMs. (A description of the corresponding
patch-clamp experiments can be found in \cite{bib9, bib5}).

Figure \ref{fi7}b depicts the spike of membrane potential produced
by two interacting EPMMs representing ensembles of sodium and
potassium channels ($N\rightarrow\infty $).

In this simulation, the sodium and potassium channels were
represented as five-state PMMs mentioned in Section \ref{sec3}.
The specific values of parameters are not important for the
purpose of this illustration.

\section{EPMM as a Distributed State Machine (DSM) } \label{sec8}
\begin{figure}[b!]
\begin{center}
 \includegraphics[width=3.0in]{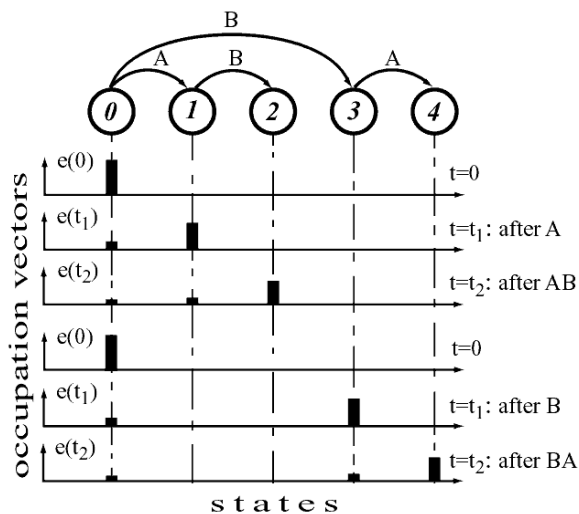}
 \end{center}
 \caption{EPMM as a distributed state machine (DSM)}
 \end{figure} \label{fi8}
Let the number of PMMs go to infinity ($N\rightarrow\infty$). In
this case EPMM is a deterministic system described by the set of
differential equations \ref{eq14} and \ref{eq15}. In some cases of
highly nonlinear input-dependent coefficients $\alpha(x,i,j)$, it
is convenient to think about this dynamical system as a
\emph{distributed state machine} (DSM). Such machine
simultaneously occupies all its discrete states, with the levels
of occupation described by the \emph{occupation vector}
$e=(e_0,... e_{n-1})$.  We replaced $\overline{e}_i$ by $e_i$,
since $N\rightarrow\infty$.

This interpretation offers a convenient language for representing
dynamical processes whose outcome depends on the sequence of input
events. In the same way as a traditional state machine is used as
a logic sequencer, a DSM can be used as an analog sequencer. The
example shown in Figure \ref{fi8} illustrates this interesting
possibility.

If the sequence of input events is $AB$ the DSM ends up "almost
completely" in state 2 (lines 1-3). The BA sequence leads to state
4 (lines 4-6). Many different implementations of a DSM  producing
this sequencing effect can be found. Here is an example of an EPMM
implementation:

\noindent Let $x=(x_1,x_2)$, $ s \in \{0,1,2,3,4\}$, and let
$\alpha(x,i,j)$ be described as follows: if input satisfies
condition $x_1>x_{thr1}\ \&\ x_2\leq x_{thr2} $ (event A) then
$\alpha(x,i,j)>0$ for transitions $(i,j) \in \{ (1,0),(4,3) \}$;
if input satisfies condition $ x_1\leq x_{thr1} \ \&\ x_2>x_{thr2}
$ (event B) then $\alpha(x,i,j)>0$ for transitions $(i,j) \in \{
(2,1),(3,0) \}$. In all other cases $\alpha(x,i,j)=0$.

This example can be interpreted as follows. If input $x_1$ exceeds
its threshold level $x_{thr1}$ before input $x_2$ exceeds its
threshold level $x_{thr2}$, the EPMM ends up "mostly" in state 2.
If these events occur in the reverse direction, the EPMM ends up
"mostly" in state 4.
\section{Why Does the Brain Need Statistical Molecular Computations?} \label{sec10}
Starting with the classical work of McCulloch and Pitts
\cite{bib11} it is well known that any computable function can be
implemented as a network of rather simple artificial neurons.
Though the original concept of the McCullough-Pitts logic neuron
is now replaced by a more sophisticated model of a leaky
integrate-and-fire (LIF) neuron \cite{bib12}, the latter model is
still very simple as compared to the EPMM formalism discussed in
the present paper.

\emph{Why does the brain need statistical molecular computations?}
\emph{Why is it not sufficient to do collective statistical
computations at the level of neural networks?} The answer to this
question is straightforward. There is not enough neurons in the
brain to implement the required computations -- such as those
associated with different effects of neuromodulation, working
memory and temporal context \cite{bib2,bib10} -- in the networks
built from the traditional artificial neurons. (Visit
\emph{www.brain0.com} to find a discussion of this critically
important issue.)
 \section{Summary} \label{sec11} 
 \begin{enumerate}
\setcounter{enumi}{0}
 \item A class of  statistical analog
computers built from large numbers of microscopic probabilistic
machines is introduced. The class is based on the abstract
computational model called \emph{Protein Molecule Machine (PMM)}.
The discussed statistical computers are represented as
\emph{Ensembles of PMMs (EPMMs)}. (Sections \ref{sec2} and
\ref{sec4}.)

\item It is postulated that at the level of neural computations
some protein molecules (e.g., ion channels) can be treated as
PMMs. That is, at this level,  specific biophysical and
biochemical mechanisms are important only as tools for the
physical implementation of PMMs with required abstract
computational properties. (Section \ref{sec3}.)

\item The macroscopic states of analog memory of the discussed
statistical computers are represented by the average relative
occupation numbers of the microscopic states of PMMs. It was
proposed \cite{bib2, bib10, bib3} that such states of cellular
analog memory are responsible for the psychological phenomena of
working memory and temporal context (mental set). (Section
\ref{sec4}.)

\item In some cases, it is useful to think of an EPMM as a
distributed state machine (DSM) that simultaneously occupies all
its discrete states with different levels of occupation. This
approach offers a convenient language for representing dynamical
processes whose outcome depends on the sequence of input events.
(Section \ref{sec8}.)

\item A computer program employing the discussed formalism was
developed. The program, called CHANNELS, allows the user to
simulate the dynamics of a cell with up to ten different
voltage-gated channels, each channel having up to eighteen states.
Two simulation modes are supported: the Monte-Carlo mode (for the
number of molecules from 1 to 10000), and the continuous mode (for
the infinite number of molecules). New software capable of
handling more complex types of PMMs is under development. (Visit
the web site www.brain0.com for more information.)
\end{enumerate}

\noindent \textbf{Acknowledgements}\newline

\noindent I express my gratitude to Prof. B. Widrow, Prof. L.
Stark, Prof. Y. Eliashberg, Prof. M. Gromov, Dr. I. Sobel, and Dr.
P. Rovner for stimulating discussions. I am especially thankful to
my wife A. Eliashberg for constant support and technical help.


\begin{thebibliography}{12}
\bibitem{bib4} Changeux, F. (1993). Chemical Signaling in the Brain. \emph{Scientific
American, November}, 58-62.
\bibitem{bib2} Eliashberg, V. (1989). Context-sensitive associative memory:
"Residual excitation" in neural networks as the mechanism of STM
and mental set. \emph{Proceedings of IJCNN-89, June 18-22, 1989,
Washington, D.C. vol. I}, 67-75. Eliashberg, V. (1989).
\bibitem{bib10}
Eliashberg, V. (1990). Universal learning neurocomputers.
\emph{Proceeding of the Fourth Annual parallel processing
symposium. California state university, Fullerton. April 4-6,
1990.} 181-191.
 \bibitem{bib3} Eliashberg, V. (1990). Molecular dynamics of short-term memory.
\emph{Mathematical and Computer modeling in Science and
Technology. vol. 14}, 295-299.
 \bibitem{bib5} Hille, B. (2001). Ion Channels of Excitable Membranes. \emph{Sinauer
Associates. Sunderland, MA}
\bibitem{bib1} Hodgkin, A.L., Huxley, A.F. 1952. A Quantitative
Description of Membrane Current and its Application to Conduction
and Excitation in Nerve. \emph {Journal of Physiology, 117},
500-544.
\bibitem{bib6}Kandel, E.R., and Spencer, W.A. (1968). Cellular
Neurophysiological Approaches in the Study of Learning.
Physiological Rev. 48, 65-134.
\bibitem{bib7} Kandel, E., Jessel,T., Schwartz, J. (2000). Principles of Neural Science.
\emph{McGraw-Hill}.
\bibitem{bib8} Marder, E., Thirumalai, V.
(2002). Cellular, synaptic and network effects of neuromodulation.
\emph{ Neural Networks 15, 479-493 }.
\bibitem{bib11} McCulloch, W. S. and Pitts, W. H. (1943).
A logical calculus of the ideas immanent in nervous activity.
\emph{Bulletin of Mathematical Biophysics, 5:115-133}.
\bibitem{bib9} Nichols, J.G., Martin, A.R., Wallace B.G., (1992)
From Neuron to Brain, \emph{Third Edition, Sinauer Associates}.
\bibitem{bib12} Spiking Neurons in Neuroscience and Technology.
\emph{ 2001 Special Issue, Neural Networks Vol. 14}.

\end{thebibliography}
\end{document}